\documentstyle[12pt,epsf,epsfig]{article}
\newcount\timecount
\newcount\hours \newcount\minutes  \newcount\temp \newcount\pmhours

\hours = \time
\divide\hours by 60
\temp = \hours
\multiply\temp by 60
\minutes = \time
\advance\minutes by -\temp
\def\hour{\the\hours}
\def\minute{\ifnum\minutes<10 0\the\minutes
            \else\the\minutes\fi}
\def\clock{
\ifnum\hours=0 12:\minute\ AM
\else\ifnum\hours<12 \hour:\minute\ AM
      \else\ifnum\hours=12 12:\minute\ PM
            \else\ifnum\hours>12
                 \pmhours=\hours
                 \advance\pmhours by -12
                 \the\pmhours:\minute\ PM
                 \fi
            \fi
      \fi
\fi
}

\def\monthname{\relax\ifcase\month 0/\or January\or February\or
   March\or April\or May\or June\or July\or August\or September\or
   October\or November\or December\else\number\month/\fi}

\def\bold#1{\setbox0=\hbox{$#1$}%
     \kern-.025em\copy0\kern-\wd0
     \kern.05em\copy0\kern-\wd0
     \kern-.025em\raise.0433em\box0 }

\def\ga{\mathrel{\raise.3ex\hbox{$>$\kern-.75em\lower1ex\hbox{$\sim$}}}}
\def\la{\mathrel{\raise.3ex\hbox{$<$\kern-.75em\lower1ex\hbox{$\sim$}}}}
\def\gev{{\rm \, Ge\kern-0.125em V}}
\def\tev{{\rm \, Te\kern-0.125em V}}
\def\beq{\begin{equation}}
\def\eeq{\end{equation}}

\def\m12{m_{1\!/2}}

\begin{document}
\begin{titlepage}
\pagestyle{empty}
\baselineskip=21pt
\rightline{hep-ph/9912324}
\rightline{CERN--TH/99-386}
\rightline{UMN--TH--1833/99}
\rightline{TPI--MINN--99/61}
\vskip 0.2in
\begin{center}
{\large{\bf Supersymmetric Dark Matter and the Energy of a Linear
Electron-Positron Collider}}
\end{center}
\begin{center}
\vskip 0.2in
{{\bf John Ellis}$^1$, {\bf Gerardo Ganis}$^2$ and {\bf Keith
A.~Olive}$^3$}\\
\vskip 0.1in
{\it
$^1${TH Division, CERN, Geneva, Switzerland}\\
$^2${Max-Planck-Institut f\"ur Physik, Munich, Germany}\\
$^3${School of Physics and Astronomy,
University of Minnesota, Minneapolis, MN 55455, USA}}\\
\vskip 0.2in
{\bf Abstract}
\end{center}
\baselineskip=18pt \noindent
We suggest that supersymmetric dark matter be used to set the
energy scale of a linear $e^+ e^-$ collider. Assuming that the
lightest supersymmetric particle (LSP) is a stable neutralino $\chi$, as
in many incarnations of the MSSM with conserved $R$ parity, previous 
calculations that include coannihilation effects
have delineated the region
of the $(m_{1/2}, m_0)$ plane where the LSP cosmological relic density 
lies in the preferred range $0.1 \la \Omega_{\chi} h^2 \la 0.3$.
We evaluate here the total cross section for $e^+ e^- \rightarrow$
visible pairs of supersymmetric particles, for different values
of $m_{1/2}$ and $m_0$, and investigate how much of the dark matter region
can be explored by $e^+ e^-$ colliders with different centre-of-mass
energies $E_{CM}$. We find that a collider with $E_{CM} = 500$~GeV
or 1~TeV
can only explore part of the cosmological region, and that a collider with
$E_{CM} = 1.5$~TeV with sufficient luminosity can explore all of
the supersymmetric dark matter region.
\vfill
\leftline{CERN--TH/99-386}
\leftline{December 1999}
\end{titlepage}
\baselineskip=18pt
\section{Introduction}

One of the most promising ideas for a high-energy accelerator
to complement the LHC is a linear $e^+ e^-$ collider (LC) with a
centre-of-mass energy $E_{CM}$ in the TeV range~\cite{LC}. The crucial
parameters of
such a LC are $E_{CM}$ and the luminosity. The optimal choice of
$E_{CM}$ is constrained by technology and cost, but should be driven
by physics arguments based on the accessibility of physics thresholds.
One established threshold in the energy range of interest is that for
$e^+ e^- \rightarrow {\bar t} t$ at about 350~GeV~\cite{top}. A second
threshold likely to be in this energy range is that for Higgs
boson ($H$) production via the reaction $e^+ e^-
\rightarrow H + Z$~\cite{ZH}. For some years~\cite{EFL}, the precision
electroweak data have
favoured a relatively light Higgs boson, as suggested independently
by supersymmetry. The most recent indication is that $M_H \la
200$~GeV
at the 95~\% confidence level~\cite{EWWG}, corresponding to a $H + Z$
threshold below about 300~GeV.

Since supersymmetry is widely considered to be one of the most promising
possible low-energy extensions of the Standard Model, it is
desirable that any new collider offer good prospects of detecting at least
some supersymmetric particles, as is the case of the LHC~\cite{LHCsusy}.
The physics argument that has usually
been employed to estimate the sparticle mass scale $\tilde m$ has been
that of
the naturalness of the gauge hierarchy, which suggests that $\tilde m
\la 1$~TeV~\cite{natural}. A supporting argument has been the concordance
of
the gauge couplings measured at LEP and elsewhere with the predictions
of supersymmetric Grand Unified Theories (GUTs)~\cite{GUTs}. However, this
argument is sensitive only logarithmically to $\tilde m$, and is also
vulnerable to GUT threshold effects due to particles beyond the minimal
supersymmetric extension of the Standard Model (MSSM). The agreement of
the Higgs mass range favoured by the precision electroweak data with
that calculated in the MSSM~\cite{MSSMHiggs} is also encouraging, but is
again only
logarithmically sensitive to $\tilde m$, and hence unable to specify it
with any accuracy.

An independent argument for new physics around the TeV scale is
provided by calculations of cold dark matter, which
yield naturally a freeze-out density in the cosmologically allowed
range: $\Omega_{CDM} h^2 \la 0.3$ (where $\Omega \equiv \rho / \rho_c$,
the
critical density, and $h$ is the Hubble expansion rate in units of
100~km/s/Mpc), and that preferred by theories of structure formation:
$0.1 \la \Omega_{CDM} h^2$, if the mass of the cold dark matter particle
is $\la 10$~TeV~\cite{Dim}. The upper limit on $\Omega_{CDM} h^2$ is fixed
by the age of
the Universe. For $\Omega_{tot} \le 1$, a lower limit on the age of the
Universe of 12~Gyr implies an upper limit $\Omega_m h^2 < 0.3$
on the total matter density, and hence $\Omega_{CDM} < \Omega_m$.
This argument does not rely on the high-redshift supernova
observations~\cite{snovae}, but they do support it.

A serendipitous prediction of
$\Omega_{CDM}$ is provided by the MSSM with $R$ parity
conservation~\cite{EHNOS}, if the lightest supersymmetric particle (LSP)
is the lightest neutralino
$\chi$, as in many versions of the MSSM. 
Indeed, it has been shown~\cite{CEOP} that the most
`natural' choices of MSSM parameters, from the point of view of the
gauge hierarchy, yield a relic LSP density in the astrophysical and
cosmological region $0.1 \la \Omega_\chi h^2 \la 0.3$.
In this case, detailed calculations of
the relic LSP abundance yield $\Omega_{CDM} \le 0.3$ only for
$m_\chi \la 600$~GeV~\cite{EFOSi}. An essential role in this
relic density calculation is played by
$\chi - \tilde \ell$ coannihilation
effects when the LSP is mainly a gaugino, which increase significantly
the upper limit on the LSP mass quoted previously~\cite{oldlimit}.

The idea we propose in this paper is that the relic density
calculation be used to specify the likelihood that a LC with
given $E_{CM}$ will be above the sparticle pair-production
threshold, and able to detect at least some supersymmetric
cross section. The answer is necessarily higher than $E_{CM} = 2 \times
m^{max}_\chi$, since the process $e^+ e^- \rightarrow \chi \chi$
is not directly observable in models with a stable neutralino LSP $\chi$.
On the other hand, as we discuss in more detail below,
$m^{max}_\chi \sim 600$~GeV is attained when
$m_\chi \sim m_{\tilde \tau}$, with $m_{\tilde \mu}, m_{\tilde e}$
not much heavier, so one might expect that
a LC with $E_{CM}$ not far above 1200~GeV
should be sufficient. As we show in more detail below, a LC with
$E_{CM} = 500$~GeV or 1~TeV would only be able to detect supersymmetry in
a fraction of the preferred dark matter region of MSSM parameter
space. A LC with $E_{CM} = 1.5$~TeV would probably cover the
preferred region, but might miss some part of the
$\chi - \tilde \ell$ coannihilation `tail' at large $m_{1/2}$,
depending on the luminosity it attains.
A LC with $E_{CM} = 2$~TeV would, on the other hand, be able to
cover all the cosmological region with a comfortable safety margin
in terms of cross section, kinematic acceptance and astrophysical
uncertainties.

\section{Summary of LSP Density Calculations}

We assume $R$ parity is conserved, otherwise there would be no
stable supersymmetric dark matter to interest us.
We work within the constrained MSSM, in which all the
supersymmetry-breaking soft scalar masses are assumed to be
universal at the GUT scale with a common value $m_0$, and the
gaugino masses are likewise assumed to be universal with 
common value $m_{1/2}$ at the GUT scale~\cite{CMSSM}. The constrained MSSM
parameters are chosen so as to yield a consistent electroweak
vacuum with a value of $\tan\beta$ that is left free. 
The LEP lower limits on MSSM particles, including the lightest
Higgs boson, suggest that $\tan\beta \ga 3$, so we consider this and
the higher value $\tan\beta = 10$. We consider two possible
values of the trilinear soft supersymmetry-breaking parameter:
$A = 0, - m_{1/2}$, the latter being the value for which
the constraint that the lowest-energy state not break charge
and colour (CCB) is weakest~\cite{CCB}, consistent with parameter
choices out to the point at the tip of the cosmological region.

When calculating the relic density of LSPs $\chi$, it is assumed
that they were in thermal equilibrium prior to freeze-out at
some temperature $T_f$.
The relic density after freeze-out is then determined by the
competition between the expansion rate of the Universe and the neutralino 
annihilation rate. Ultimately, the relic density is inversely related to
the effective annihilation cross section
$\sigma_{eff}$, which falls off as the square of the supersymmetry 
breaking scale.
Thus, as the supersymmetry breaking scale is increased, the annihilation
cross section decreases and the relic density increases. 
This is why an upper limit
to the relic density puts an upper limit on the sparticle mass scale,
and on the mass of the neutralino LSP, in particular. In regions where the
neutralino is
mainly a gaugino (usually a bino), as in many models of interest,
such as those with GUT-scale universality relations
among the sparticle masses, the annihilation rate is dominated by
sfermion exchange.  As one approaches the upper limit on the neutralino
mass, the cross section is maximized by taking sfermion masses as small
as possible: in this case, the sleptons $\tilde \ell$ are nearly
degenerate with
the neutralino LSP $\chi$~\footnote{The GUT universality conditions then
imply that the squarks are considerably heavier.}. 

When the LSP is nearly degenerate with the next-to-lightest
supersymmetric particle (NLSP), it is known~\cite{oldcoann}
that new important coannihilation channels must be included to
determine the relic neutralino density.  Thus, in addition to
the self-annihilation process
$\chi \chi
\rightarrow$ anything, the effective annihilation cross section
includes important contributions from coannihilation processes involving
slightly heavier sparticles
$\tilde X,
\tilde Y$:
$\chi \tilde X
\rightarrow$ anything,
$\tilde X \tilde Y
\rightarrow$ anything, weighted by the corresponding
Boltzmann density suppression factors:
\begin{equation}
\sigma_{eff} \sim \sigma(\chi \chi) + \Sigma_{\tilde X}
e^{- (m_{\tilde X} - m_\chi) / T_f} \sigma (\chi \tilde X)
+ \Sigma_{\tilde X, \tilde Y} e^{- (m_{\tilde X} + m_{\tilde Y} - 2 
m_\chi) / T_f} \sigma (\tilde X \tilde Y)
\label{effsigma}
\end{equation}
In the parameter region of interest after 
taking into account the LEP exclusions of light sparticles,
the most important coannihilation processes are those
involving the NLSP $\tilde \tau$ and other sleptons: $\tilde e,
\tilde \mu$, which are all taken into account in the following
analysis~\cite{EFOSi}. Several of these coannihilation cross sections are
much larger than that for $\chi \chi$ annihilation close to threshold,
because they do not exhibit $P$-wave suppressions. Therefore,
coannihilation is an essential complication.

As noted above, since the resulting LSP relic density $\Omega_\chi h^2$
increases as
$\sigma_{eff}$ decreases, and since 
$\sigma_{eff}$ decreases as $m_0, m_{1/2}$ increase, one
expects generically that $\Omega_\chi h^2$ should {\it increase} with
{\it increasing}
$m_0, m_{1/2}$. This simple correlation is complicated in the
presence of nearby $s$-channel $Z^0$ and Higgs poles in
the annihilation cross sections, but the LEP exclusions now
essentially rule out this possibility~\cite{EFGOS}.
As mentioned earlier in the paper, the preferred range of cold dark matter
density is $0.1 \la \Omega_{CDM} h^2 \la 0.3$. It is possible that
all the cold dark matter may not consist of LSPs $\chi$, so we can at
best assume that $\Omega_{\chi} h^2 \le \Omega_{CDM} h^2 \la 0.3$.
However, this {\it upper limit} on $\Omega_\chi h^2$ is sufficient to
infer an {\it upper limit} on $m_0, m_{1/2}$~\footnote{On the other hand,
the {\it lower bound} on $\Omega_{CDM} h^2 \ga 0.1$ cannot be
transferred to a lower bound on $\Omega_{\chi}$, and hence there are
{\it no corresponding lower bounds} on $m_0, m_{1/2}$, except for those
imposed by slepton searches and/or the requirement that the
$\tilde \tau$ not be the LSP.}. In
\cite{EFOSi}, the values of the two key supersymmetry-breaking inputs
$m_0, m_{1/2}$ were constrained
so that neutralino relic density should fall within the desired range.
Roughly speaking, when $m_{1/2} \la 400$ GeV, there is a relatively
broad allowed range for $m_0$ between about 50 and 150~GeV, depending on
$\tan \beta, A$ and the sign of $\mu$.  For values of $m_{1/2} \ga 400$
GeV,
coannihilation becomes important, and $m_0$ is restricted to a relatively
narrow range of typical thickness $\delta m_0 \sim 20$~GeV. The maximum
value of $m_{1/2}$ is
determined by the point where there is no longer any value of $m_0$, such
that the
neutralino mass is less than the $\tilde \tau_R$ mass and $\Omega_{CDM} 
h^2 <
0.3$. This occurs when $m_{1/2} \simeq 1400$ GeV, corresponding to the
neutralino mass of about 600 GeV mentioned previously.

{\it This is the essence of our argument that the relic density
calculation can be used to specify the $e^+ e^-$ collider energy
required to produce sparticles.}

The upper limit to the neutralino mass including coannihilation effects
of $m_\chi \la 600$~GeV is relatively insensitive to such MSSM parameters
as
$\tan\beta$ and $A$. As in~\cite{EFOSi}, we consider here the two
cases $\tan\beta = 3, 10$, and initially set $A$ close to the weak-CCB
value $A = - m_{1/2}$. As mentioned earlier,
the upper limit on $m_\chi$ implies that the threshold for 
pair-producing sparticles must be at least $E_{CM} = 1200$~GeV.
In fact, when the limit $m_\chi \sim 600$~GeV is reached, one also
has $m_\chi = m_{\tilde \tau_1}$, where the NLSP $\tilde \tau_1$
is the lighter stau mass eigenstate, so the threshold
for the reaction 
$e^+ e^- \rightarrow {\tilde \tau}^+ {\tilde \tau}^-$
is also $\sim 1200$~GeV. Moreover, the mass of the ${\tilde e}_R$ is
also not far above $600$~GeV, so the threshold for $e^+ e^- 
\rightarrow {\tilde e}_R^+ {\tilde e}_R^-$ is also not far beyond
$1200$~GeV. In addition, it is easy to check that even if one
allows $m_\chi < m_{\tilde \tau_1}$, which is possible if
$m_\chi < 600$~GeV, the threshold for 
$e^+ e^- \rightarrow {\tilde \tau}^+ {\tilde \tau}^-$
is never above $1200$~GeV. These arguments are all suggestive
that $E_{CM} = 1200$~GeV may be sufficient for an $e^+ e^-$
linear collider to observe supersymmetry, but any such conclusion
must hinge upon the analysis of the observability of the
sparticle pair-production cross section that we undertake next.

\section{Analysis of Sparticle Pair-Production Cross Sections}

In order to determine the region of the $(m_0, m_{1/2})$ plane
that can be explored with a linear $e^+ e^-$ collider of given $E_{CM}$,
we have calculated the total observable production cross section for the 
pair production of sparticles $e^+ e^- \rightarrow {\tilde X}
{\tilde Y}$,
where $\tilde X$ and $\tilde Y$ are not necessarily particle and
antiparticle~\cite{package}. In this context, `observable' means that we
do not
include pair production of the LSP: $e^+ e^- \rightarrow \chi \chi$.
Nor do we include sneutrino pair production: $e^+ e^- \rightarrow {\tilde
\nu} {\tilde {\bar \nu}}$, although some $\tilde \nu$ decays
might be visible. Also, the production cross sections for heavier
neutralinos $\chi'$, e.g., $e^+ e^- \rightarrow \chi \chi'$, are corrected
for invisible $\chi'$ decay branching ratios. Finally, we assume that
the ordinary particles emitted in a sparticle decay chain are
observable only if the mass difference $\Delta M > 3$~GeV.

We assume an integrated luminosity ${\cal L} = 100$~fb$^{-1}$~\cite{LC}.
In order to estimate the corresponding sensitivity to the new-physics
cross section $\sigma$,
the relevant quantity is $B \equiv \sqrt{\sigma_{bg}}/ \epsilon$,
where $\sigma_{bg}$ is the residual cross section for 
background processes, and $\epsilon$ is the signal-detection efficiency.
As usual, a five-standard-deviation discovery is likely if   
$\sigma > 5 \times B/\sqrt{{\cal L}}$, whereas, in the absence of any 
observation, new-physics processes with $\sigma > 2 \times B/\sqrt{{\cal L}}$
will be excluded at about the 95~\% confidence level. 
At LEP~2, for mass differences between the produced sparticle and the LSP 
that are
not too small, the background to searches for charginos $\chi^{\pm}$ 
and sleptons is mainly due to $W^{\pm}$ production, and typical values 
for $B$ were in the range $3\div 6$~(fb$^{+\frac{1}{2}}$). At the LC
we expect cleaner background conditions for both slepton and chargino 
searches because the $W^{\pm}$ should be more easily distinguishable,
and also $\sigma(e^+ e^- \rightarrow W^+ W^-)$ is smaller. 
It is therefore likely that B is smaller than at LEP~2. 
We adopt a conservative approach and
scale $B$ roughly by $\sigma(e^+ e^- \rightarrow W^+ W^-)$,
taking $B = 2$~(fb$^{+\frac{1}{2}}$), which
gives a lower limit on the discoverable cross
section of 1~fb, and an exclusion upper limit of 0.4~fb.

Fig.~1 shows the physics discovery reach in the $(m_0, m_{1/2})$ plane 
for $\tan\beta = 3, 10$
provided by the processes $e^+e^- \rightarrow {\tilde \ell}^+ {\tilde
\ell}^-$, neutralinos and charginos ${\chi}^+ {\chi}^-$ for 
collisions at $e^+ e^-$ collisions at $E_{CM} = 500,
1000, 1250, 1500$~GeV, compared with the allowed cold dark matter
region (shaded). 
\begin{figure}
\begin{center}
\mbox{\epsfig{file=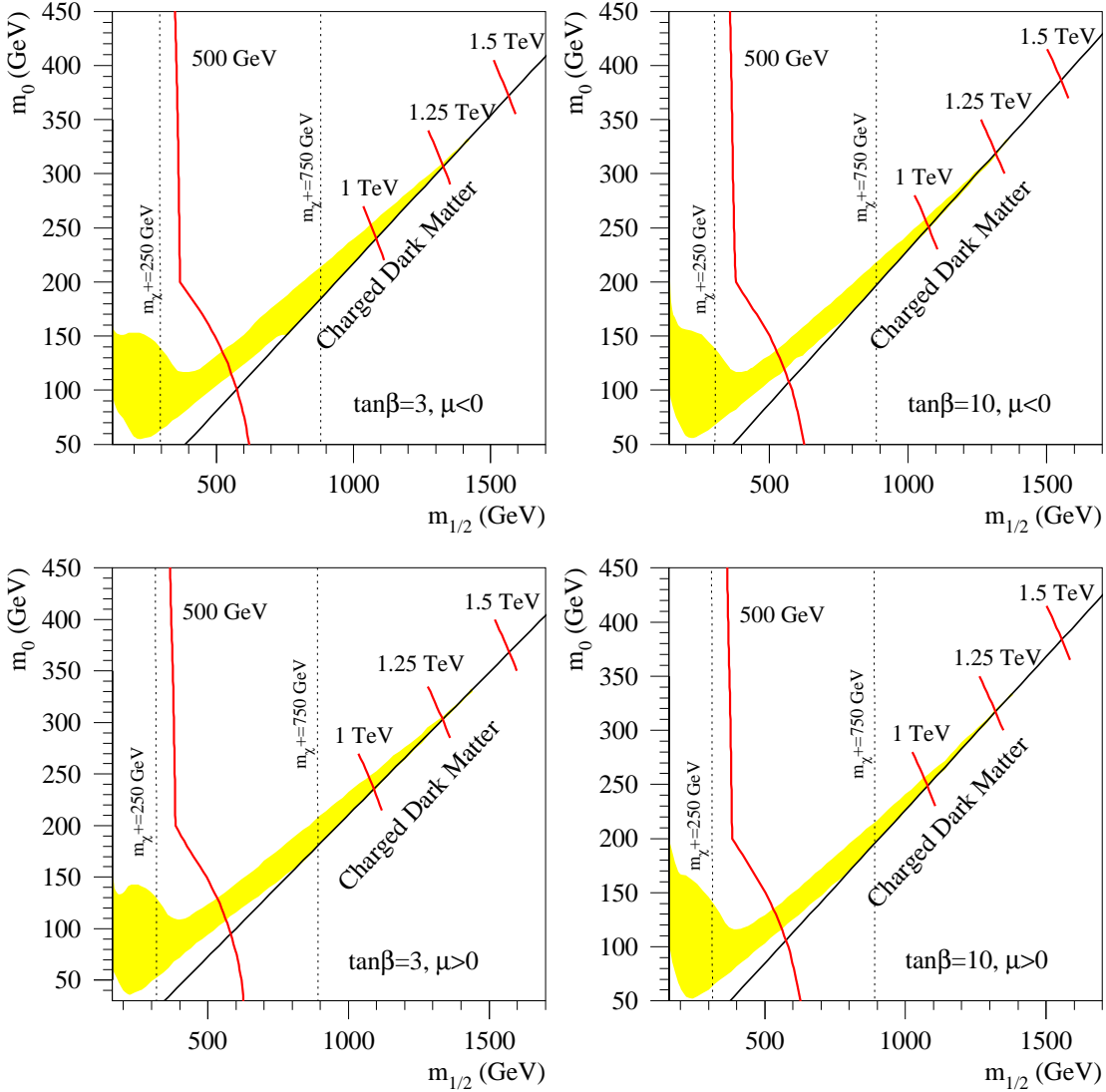,height=16.5cm}}
\end{center}
\caption[.]{\it
Discovery sensitivity in the $(m_0, m_{1/2})$ plane for $\tan\beta =
3$ (left panels) and $10$ (right panels) provided by searches for 
$e^+ e^- \rightarrow {\tilde \ell}^+ {\tilde \ell}^-$ 
and neutralinos (solid lines) and ${\chi}^+
{\chi}^-$ (broken lines) for collisions at $E_{CM}
= 500, 1000, 1250, 1500$~GeV. The allowed cold dark matter regions are
shaded. The top (bottom) panels are for $\mu < (>)
0$, and the value $A = - m_{1/2}$ is used.}
\end{figure}
The solid lines in Fig.~1 correspond to the estimated discovery
cross section of
1~fb for $e^+ e^- \rightarrow {\tilde \ell}^+ {\tilde \ell}^-$, and the
broken
lines to the kinematic limit $m_{\chi^\pm} = E_{CM} / 2$. 
We see no big differences between the plots for the different signs
of $\mu$, nor indeed for the different values of $\tan\beta$.
We note that
$e^+ e^- \rightarrow {\tilde \ell}^+ {\tilde \ell}^-$ (solid lines)
provides
the greatest reach for each of the values $E_{CM} = 500, 1000, 1250,
1500$~GeV studied, and that chargino pair production $e^+ e^- \rightarrow
{\chi}^+ {\chi}^-$ (broken lines) becomes progressively less important as
$E_{CM}$ increases.

We see in Fig.~1 the extent to which the region favoured by the
cosmological
requirement that $0.1 \le \Omega_\chi h^2 \le 0.3$ may be covered
by LC searches at different energies. In particular,
about a half of this region is covered by sparticle searches at
$E_{CM} = 500$~GeV, a somewhat larger fraction (but not all) is
covered at $E_{CM} = 1000$~GeV, and full coverage of the favoured region
is approached only when $E_{CM} = 1500$~GeV~\footnote{We note in
passing that a LC with $E_{CM} = 500$~GeV would have seemed perfectly
adequate if coannihilation were not taken into account.}.
The reason why more than 1200~GeV
is required is the $P$-wave threshold suppression for the
observable processes with the lowest thresholds near the point
of the cosmological region, namely the reactions 
$e^+ e^- \rightarrow {\tilde \ell}_R^+ {\tilde \ell}_R^-$.

Fig.~2 shows as three-dimensional `mountains' the full
observable sparticle cross section for $\tan\beta =
10$ and $\mu > 0$ for $E_{CM} = 500, 1000, 1250$ and 1500~GeV,
including
also other pair-production processes. 
\begin{figure}
\begin{center}
\mbox{\epsfig{file=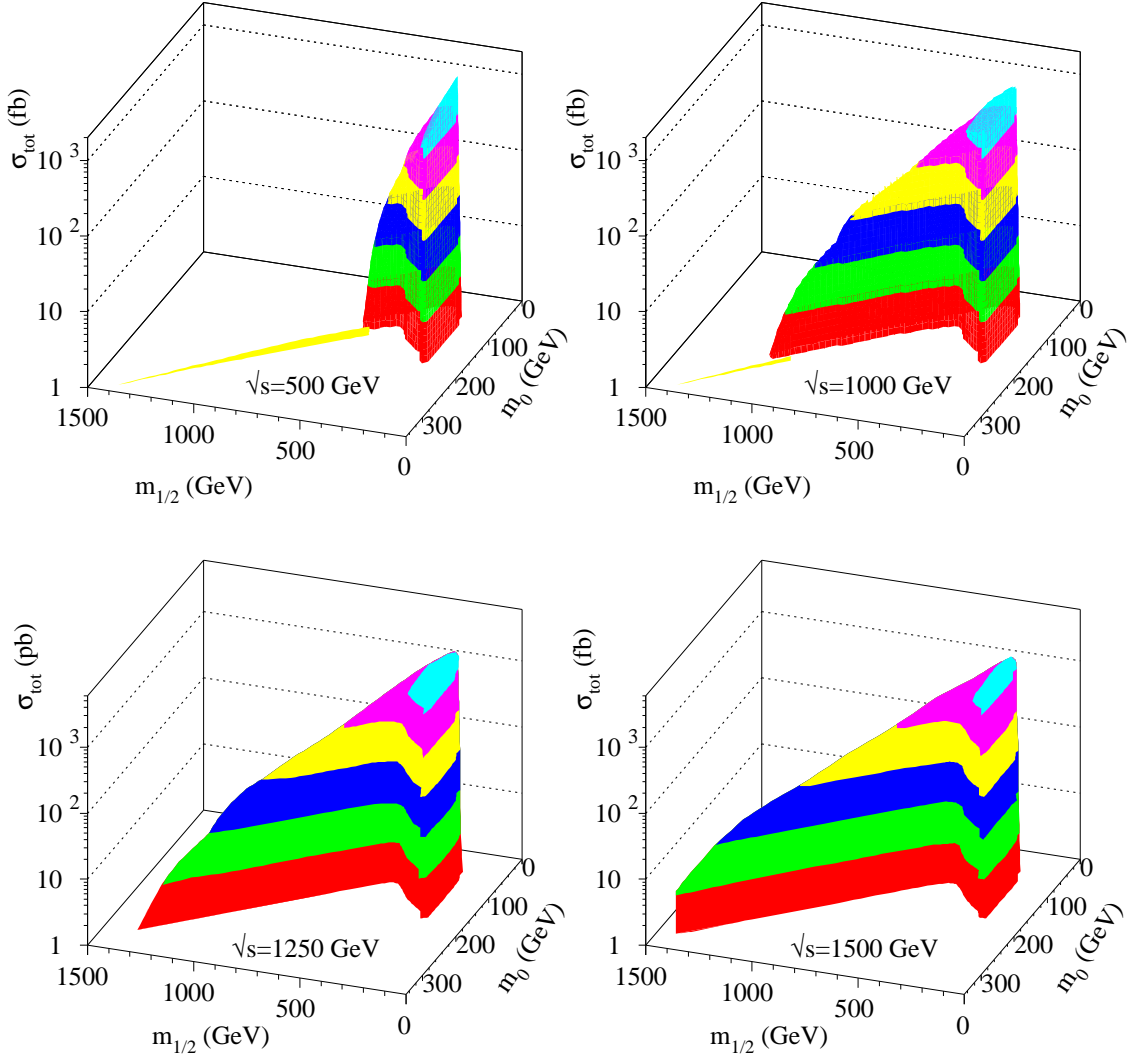,height=16.5cm}}
\end{center}
\caption[.]{\it The observable sparticle pair-production cross section
in the $(m_{1/2}, m_0)$ plane for $\tan\beta = 10$, $\mu > 0$ and
$A = -m_{1/2}$, for $E_{CM} = 500, 1000, 1250$ and $1500$~GeV. Note that
the
vertical scale is logarithmic, and that cross-section contours
are indicated by changes in the shading of the cross-section
`mountain'. The cosmologically-preferred
domain of $(m_{1/2}, m_0)$ is visible in projection in the top 
two panels: in the bottom two panels, it is
obscured by the `mountain'.}
\end{figure}
The irregularities in the outline of the three-dimensional 
`mountain' plot
correspond to the opening up of different sparticle pair-production
thresholds.
We see again that $E_{CM} = 500$~GeV
is not adequate to cover much of the cosmological region, that
$E_{CM} = 1000$~GeV does not cover a
significant fraction of the high-$m_{1/2}$ tail opened up by
coannihilation, and that $E_{CM} \ge 1500$~GeV covers 
the cosmological region.
We find similar
features 
for $\tan\beta = 10$ and $\mu < 0$, and also for $\tan\beta = 3$
and both signs of $\mu$ (not shown).

We now return to the tip of the cosmological tail,
which occurs when $m_\chi \sim 630 (610)$~GeV for $\tan\beta = 3 (10)$
for our default option $\Omega_\chi h^2 \le 0.3$,
and explore in more detail how much $E_{CM}$ beyond 1200~GeV is
required to be sure of detecting supersymmetry. Fig.~3 shows
the contributions to the effective observable cross section
from the dominant reactions $e^+ e^- \rightarrow {\tilde \ell}_{L,R}^+
{\tilde \ell}_{L,R}^-$.
\begin{figure}
\begin{center}
\mbox{\epsfig{file=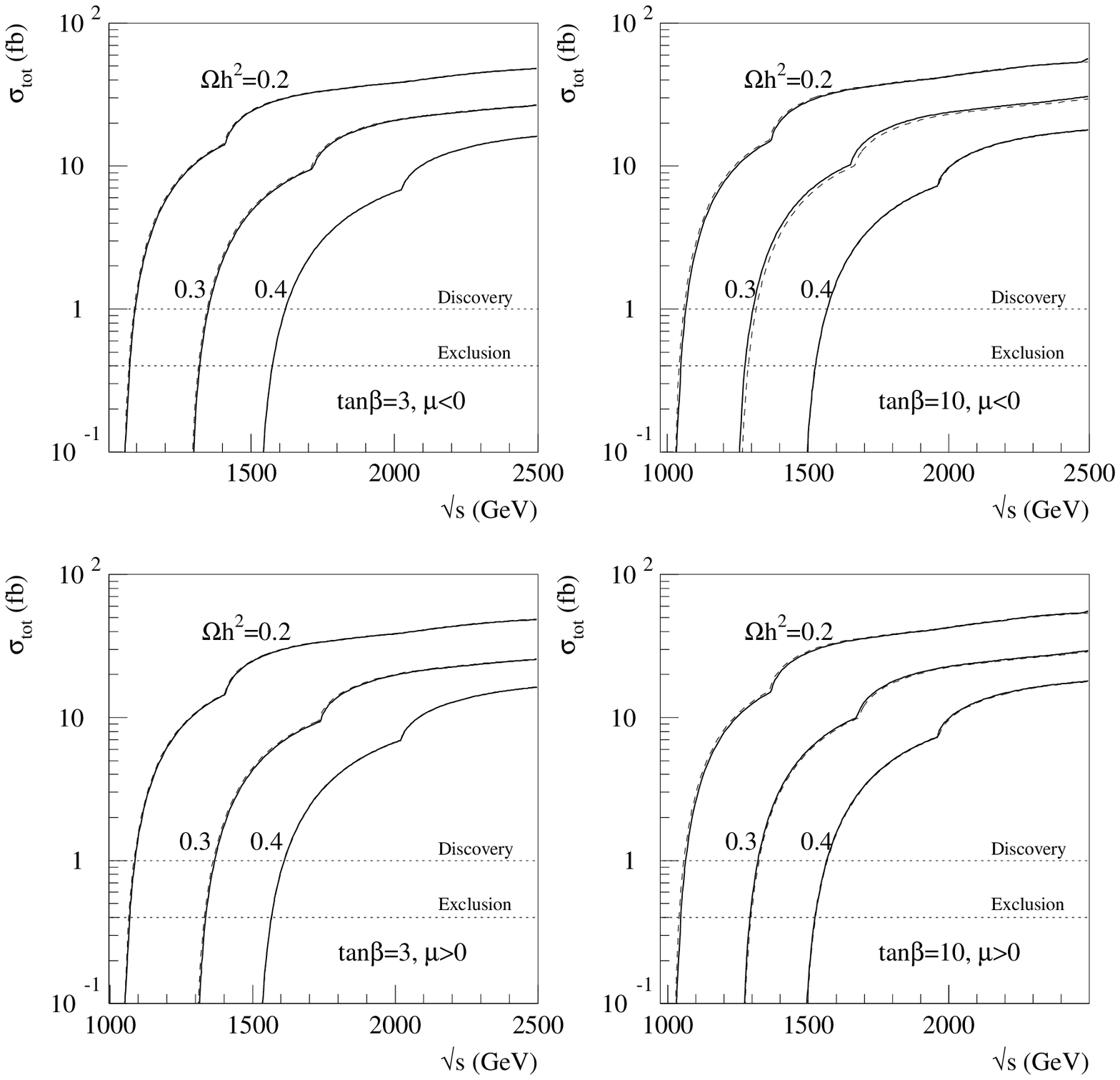,height=16.5cm}}
\end{center}
\caption[.]{\it The total observable cross section for $e^+ e^-
\rightarrow {\tilde \ell}^+ {\tilde \ell}^-$ processes, as a function
of $E_{CM}$, for the points at the tips of the region of $m_{1/2}, m_0$
allowed for $\Omega_\chi h^2 \le 0.2, 0.3$ (our preferred choice) and 
0.4, with the usual choices $\tan\beta = 3, 10$ and both signs of $\mu$.
The solid lines are for $A = - m_{1/2}$, and the dashed lines for $A = 0$.
The horizontal
broken lines are our estimates of the possible discovery and exclusion
cross sections.}
\end{figure}
Close to threshold, only pair production of the
${\tilde \ell}_R$ states is accessible, which exhibits a $P$-wave
suppression. The associated-production process
$e^+ e^- \rightarrow {\tilde e}_L {\tilde e}_R$ kicks in
at somewhat higher energies, and rapidly
dominates, because of its $S$-wave threshold.
This is the origin of the kink seen in the rise of the
total cross section in each of the panels of Fig.~3,
where the discovery and exclusion sensitivities are also shown
as horizontal broken lines.
We see that $E_{CM}$ only just
above 2$m_\chi \sim 1200$~GeV is not sufficient for sparticle
discovery, because of the small observable cross section.
We recall that, for our assumed integrated luminosity of
100~fb$^{-1}$ and detector performances, the discovery
cross-section limit would be 1~fb, as indicated by the
upper horizontal broken line in Fig.~3. Of course, this may be
altered by different assumptions on the integrated luminosity
and/or detection efficiency~\footnote{We note, in particular,
that higher luminosities may be achievable at higher $E_{CM}$.}.

Each of the panels in Fig.~3
exhibits alternative curves to be compared with our default choices
$\Omega_{\chi} h^2 = 0.3$ and $A = 0$. The curves for
$\Omega_{\chi} h^2 = 0.4$ are for instruction only. In this
case, one finds $m_\chi \la 740 (710)$ GeV for $\tan \beta = 3 (10)$,
but it is very
difficult to reconcile such a large value 
of $\Omega_\chi h^2$ with
the emerging measurements of cosmological parameters~\footnote{For
the record, for $\Omega_\chi h^2 < 0.5$,
the upper limit on the neutralino mass increases to 
$m_\chi \la 830 (800)$ GeV for $\tan \beta = 3 (10)$.}.
In fact, we actually believe that allowing $\Omega_\chi h^2 \le 0.3$
is already quite conservative. For the preferred 
observational value $h \sim
1/\sqrt{2}$, this would correspond to $\Omega_\chi \le 0.6$,
which extends far beyond the currently favoured range
$\Omega_\chi \le 0.4$. If instead one enforces 
$\Omega_\chi h^2 \le 0.2$, one finds that the
maximum value of the LSP mass becomes $m_\chi \sim 520(500)$~GeV, for
$\tan \beta = 3 (10)$ and
$E_{CM} = 1500$ TeV would be adequate, as seen in Fig.~3.
Indeed, in this case, 
$E_{CM} = 1200$~GeV would be sufficient to cover all the
region of the $(m_0, m_{1/2})$ favoured by cosmology.
We also show in Fig.~3 comparisons between the cross sections
at the extreme points for $A = 0$ and $-m_{1/2}$.
Our conclusions are
clearly insensitive to the ambiguity in the choice of $A$.

\section{Conclusions}

Finally, we show in Fig.~4 the fraction of the cosmologically-allowed
region of the $(m_{1/2}, m_0)$ plane that can be explored by a LC
as a function of the accessible limiting cross section $\sigma_{lim}$,
for different values of $E_{CM}$. When the detector perfomances are
specified, the values of $\sigma_{lim}$ correspond to different values of
the available luminosity, as indicated.
\begin{figure}
\begin{center}
\mbox{\epsfig{file=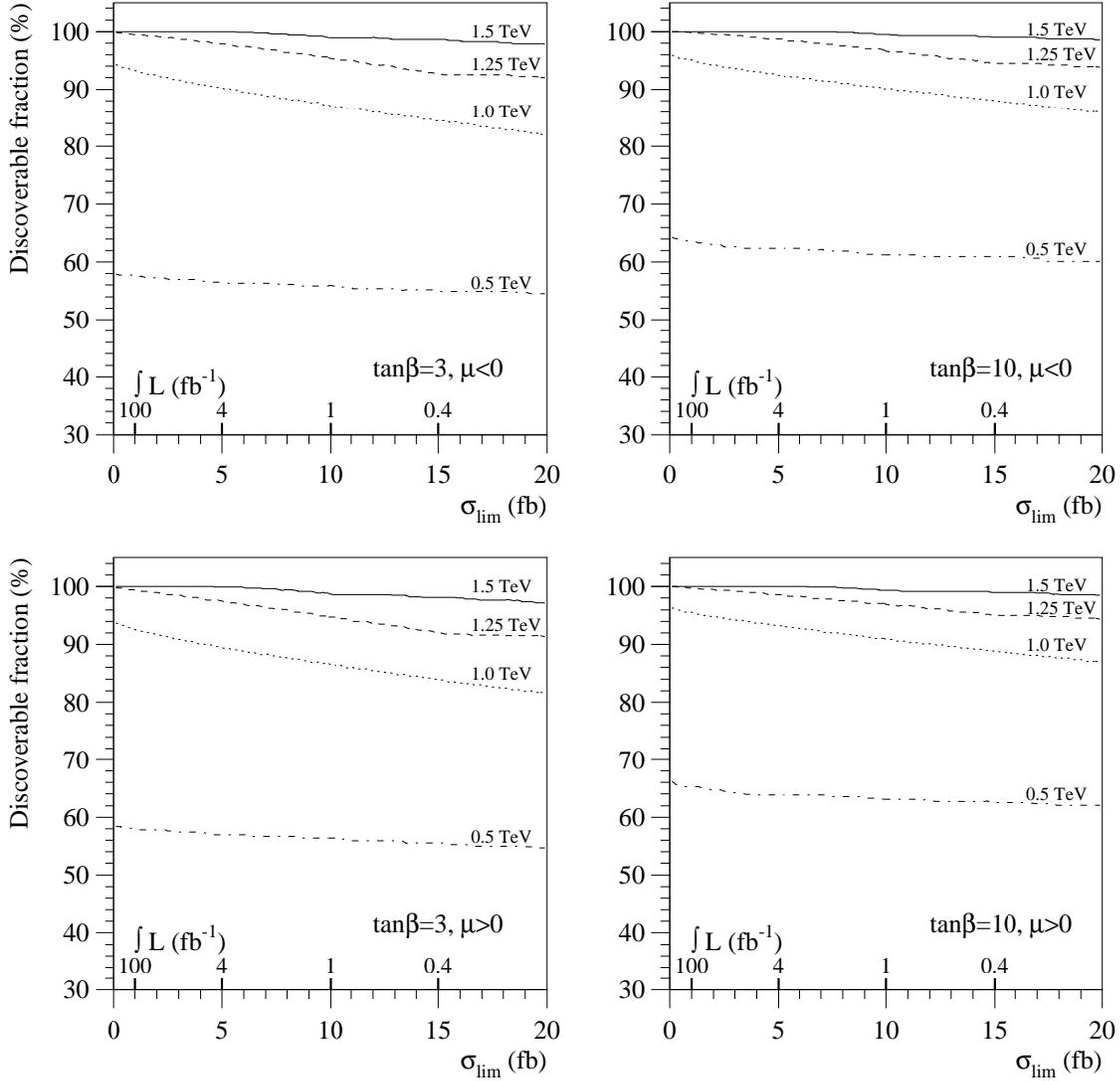,height=16.5cm}}
\end{center}
\caption[.]{\it The discoverable fractions of the
region of the $(m_{1/2}, m_0)$ plane allowed by cosmology for
$0.1 \le \Omega_\chi h^2 \le 0.3$ that is accessible to a LC
as a function of accessible cross section $\sigma_{lim}$, for
different values of $E_{CM}$, for each of our scenarios $\tan\beta
=3, 10$ and both signs of $\mu$, assuming $A = -m_{1/2}$.
Also indicated is the correspondence between luminosity and
$\sigma_{lim}$ for the detector performances assumed in this
paper. }
\end{figure}
We see in Fig.~4 that a LC with $E_{CM} = 1.5$~TeV would
cover all the
cosmological region if $\sigma_{lim} \la 5$~fb~\footnote{A LC 
with $E_{CM} = 2$~TeV would always cover all the
cosmological region, even for a very pessimistic assumption on
$\sigma_{lim}$.}, and one with $E_{CM} =
1.25$~TeV if $\sigma_{lim} \la 0.5$~fb. On the other hand, a LC
with $E_{CM} = 1$~TeV could never cover all the cosmological region,
and a LC with $E_{CM} = 0.5$~TeV covers $\sim 60$~\% of 
it~\footnote{Fig.~4 is plotted using a linear measure for the
cosmological region. The
prospects for lower-energy machines would seem brighter if one used a
logarithmic measure of the
parameter space, e.g., using this measure, a LC with $E_{CM} = 0.5$~TeV
would cover over 80~\% of the cosmological region.}. 

The conclusions to be drawn from this analysis are somewhat subjective,
since they
depend how much you are prepared to bet at what odds. It could
well be that new cosmological data might inform better your choice.
For example, you could become more sanguine about the prospects for
a lower-energy LC if the upper limit on $\Omega_\chi h^2$ could be
decreased to 0.2. Our point in this paper has been to establish
that there is a phenomenological connection between the LC energy and
supersymmetric
dark matter, and we believe that Fig.~4 summarizes the best advice we can
offer at the beginning of the third millennium.

\vskip 0.5in
\vbox{
\noindent{ {\bf Acknowledgments} } \\
\noindent 
We thank Toby Falk for many related discussions. 
The work of K.A.O. was supported in part by DOE grant
DE--FG02--94ER--40823.}


\begin{thebibliography}{99}

\bibitem{LC}
See, for example:
H.~Murayama and M.E.~Peskin,
Ann.\ Rev.\ Nucl.\ Part.\ Sci.\ {\bf 46} (1996) 533;
E.~Accomando {\it et
al.},
ECFA/DESY LC Physics Working Group Collaboration,
Phys.\ Rept.\ {\bf 299} (1998) 1.

\bibitem{top}
F.~Abe {\it et al.},
CDF Collaboration,
Phys.\ Rev.\ Lett.\ {\bf 73} (1994) 225.

\bibitem{ZH}
J. Ellis, M.K. Gaillard and D.V. Nanopoulos, Nucl.\ Phys.\ {\bf B106}
(1976) 292;
B.W. Lee, C. Quigg and H.B. Thacker, Phys.\ Rev.\ {\bf D16} (1977) 1519;
B.L. Ioffe and V.A. Khoze, Sov.\ J.\ Part.\ Nucl.\ {\bf 9} (1978) 50.

\bibitem{EFL}
J. Ellis, G.L. Fogli and E. Lisi, Phys.\ Lett.\ {\bf B318} (1993) 148.

\bibitem{EWWG} 
LEP Electroweak Working Group,
{\tt http://www.cern.ch/LEPEWWG/Welcome.html}.

\bibitem{LHCsusy}
I. Hinchliffe, F.E. Paige, M.D. Shapiro, J. Soderqvist and W. Yao,
Phys.\ Rev.\ {\bf D55} (1997) 5520;
S. Abdullin and F. Charles, Nucl.\ Phys.\ {\bf B547} (1999) 60.

\bibitem{natural}
L. Maiani, {\it Proc. Summer School on Particle Physics}, Gif-sur-Yvette,
1979 (IN2P3, Paris, 1980) p.3;
G. 't Hooft, {\it Recent Developments in Field Theory}, eds. G. 't Hooft
{\it et al.}, (Plenum Press, New York, 1980);
E. Witten, Nucl.\ Phys.\ {\bf B188} (1981) 513;
R.K. Kaul, Phys.\ Lett.\ {\bf 109B} (1982) 19.

\bibitem{GUTs}
J. Ellis, S. Kelley and D.V. Nanopoulos, Phys.\ Lett.\ {\bf B260} (1991)
131;
U. Amaldi, W. de Boer and H. Furstenau, Phys.\ Lett.\ {\bf B260} (1991)
447;
P. Langacker and M. Luo, Phys.\ Rev.\ {\bf D44} (1991) 817.

\bibitem{MSSMHiggs}
M.~Carena, S.~Heinemeyer, C.E.~Wagner and G.~Weiglein,
hep-ph/9912223; and references therein.

\bibitem{Dim}
S.~Dimopoulos,
Phys.\ Lett.\ {\bf B246} (1990) 347.

\bibitem{snovae}
A.G. Riess {\it et al.}, Astron.\ J.\ {\bf 116} (1998) 1009;
S. Perlmutter {\it et al.}, astro-ph/9812133.

\bibitem{EHNOS}
J. Ellis, J.S. Hagelin, D.V. Nanopoulos, K.A. Olive
and M. Srednicki, Nucl. Phys. {\bf B238} (1984) 453.


\bibitem{CEOP}
P.H.~Chankowski, J.~Ellis, K.A.~Olive and S.~Pokorski,
Phys.\ Lett.\ {\bf B452} (1999) 28.

\bibitem{EFOSi}
J.~Ellis, T.~Falk and K.A.~Olive,
Phys.\ Lett.\ {\bf B444} (1998) 367;
J.~Ellis, T.~Falk, K.A.~Olive and M.~Srednicki,
hep-ph/9905481.

\bibitem{oldlimit}
K.A. Olive and M. Srednicki, Phys.\ Lett.\ {\bf B230} (1989) 78 and
Nucl.\ Phys.\ {\bf B355} (1991) 208;
K. Griest, M. Kamionkowski and M.S. Turner, Phys.\ Rev.\ {\bf D41} (1990)
3565;
G. Kane, C. Kolda, L. Roszkowski and J. Wells, Phys.\ Rev.\
{\bf D49} (1994) 6173.

\bibitem{CMSSM}
For reviews, see:
H.P. Nilles, Phys. Rep. {\bf 110} (1984) 1;
H.E. Haber and G.L. Kane, Phys. Rep. {\bf 117} (1995) 75.


\bibitem{CCB}
H. Baer, M. Brhlik and D. Casta\~no, Phys.\ Rev.\ {\bf D54} (1996) 6944;
S. Abel and T. Falk, Phys.\ Lett.\ {\bf B444} (1998) 427.

\bibitem{oldcoann}
K. Griest and D. Seckel, Phys.\ Rev.\ {\bf D43} (1991) 3191;
S. Mizuta and M. Yamaguchi, Phys.\ Lett.\ {\bf B298} (1993) 120.

\bibitem{EFGOS}
J. Ellis, T. Falk, K.A. Olive and M. Schmitt,
Phys.\ Lett.\ {\bf B388} (1996) 97
and Phys.\ Lett.\ {\bf B413} (1997) 355;
J.~Ellis, T.~Falk, G.~Ganis, K.A.~Olive
and M.~Schmitt,
Phys.\ Rev.\ {\bf D58} (1998) 095002.

\bibitem{package}
G. Ganis,
{\tt http://alephwww.cern.ch/${\tilde {~}}$ganis/MSMLIB/msmlib.html}.

\end{thebibliography}
\end{document}